\documentclass[aps,pre,amsmath,10pt,twocolumn,superscriptaddress]{revtex4-1}
\usepackage[dvipdfmx]{graphicx}
\usepackage{color}
\usepackage{bm}

\begin{document}
\title{Group-theoretical high-order rotational invariants for structural representations: Application to linearized machine learning interatomic potential}
\author{Atsuto \surname{Seko}}
\email{seko@cms.mtl.kyoto-u.ac.jp}
\affiliation{Department of Materials Science and Engineering, Kyoto University, Kyoto 606-8501, Japan}
\affiliation{Center for Elements Strategy Initiative for Structure Materials (ESISM), Kyoto University, Kyoto 606-8501, Japan}
\affiliation{JST, PRESTO, Kawaguchi 332-0012, Japan}
\affiliation{Center for Materials Research by Information Integration, National Institute for Materials Science, Tsukuba 305-0047, Japan}
\author{Atsushi \surname{Togo}}
\affiliation{Center for Elements Strategy Initiative for Structure Materials (ESISM), Kyoto University, Kyoto 606-8501, Japan}
\author{Isao \surname{Tanaka}}
\affiliation{Department of Materials Science and Engineering, Kyoto University, Kyoto 606-8501, Japan}
\affiliation{Center for Elements Strategy Initiative for Structure Materials (ESISM), Kyoto University, Kyoto 606-8501, Japan}
\affiliation{Center for Materials Research by Information Integration, National Institute for Materials Science, Tsukuba 305-0047, Japan}
\affiliation{Nanostructures Research Laboratory, Japan Fine Ceramics Center, Nagoya 456-8587, Japan}

\date{\today}

\begin{abstract}
Many rotational invariants for crystal structure representations have been used to describe the structure-property relationship by machine learning.
The machine learning interatomic potential (MLIP) is one of the applications of rotational invariants, which provides the relationship between the energy and the crystal structure.
Since the MLIP requires the highest accuracy among machine learning estimations of the structure-property relationship, the enumeration of rotational invariants is useful for constructing MLIPs with the desired accuracy.
In this study, we introduce high-order linearly independent rotational invariants up to the sixth order based on spherical harmonics and apply them to linearized MLIPs for elemental aluminum.
A set of rotational invariants is derived by the general process of reducing the Kronecker products of irreducible representations (Irreps) for the SO(3) group using a group-theoretical projector method.
A high predictive power for a wide range of structures is accomplished by using high-order invariants with low-order invariants equivalent to pair and angular structural features.
\end{abstract}

\maketitle

\section{Introduction}

The machine-learning interatomic potential (MLIP) based on a large dataset generated by density functional theory (DFT) calculations is beneficial for significantly improving the accuracy and transferability of interatomic potentials\cite{
behler2007generalized,
bartok2010gaussian,
behler2011atom,
PhysRevB.90.024101,
PhysRevB.90.104108,
PhysRevB.92.054113,
PhysRevMaterials.1.063801,
Thompson2015316,
PhysRevMaterials.1.043603,
PhysRevLett.114.096405,
PhysRevB.95.214302,
doi-10.1063-1.4930541,
PhysRevB.92.045131,
QUA:QUA24836,
doi-10.1137-15M1054183}.
Applications of MLIP have been increasing for not only atomistic simulations in large systems but also global structure optimizations\cite{PhysRevLett.120.156001,podryabinkin2018accelerating,GUBAEV2019148}, which require a high predictive power over a wide range of configurations.
Similarly to conventional interatomic potentials, the MLIP is based on the common idea that the total energy of a structure may be divided into the energies of the constituent atoms of a system.
In the formulation of MLIPs, the atomic energy originating from atomic interactions with neighboring atoms is formulated as a function of a set of numerous quantities depending on its neighboring environment called structural ``features'' or ``descriptors''.
A number of models have been employed to describe a function or a mapping from structural features to the atomic energy, including an artificial neural network, a Gaussian process model, a polynomial function and a simple linear model.

Recently, many studies on estimating the structure-property or compound-property relationship by machine learning have been reported.
In such a machine learning estimation, the invariant properties of a set of target systems such as translational and rotational invariances play a key role in generating structural or compound features.
Regarding the structure-energy relationship of an MLIP, the total energy of a crystalline system has been modeled as a function of a wide range of invariant quantities, 
including order parameters depending on pairs and angles among three atoms\cite{behler2007generalized,jose2012construction,bartok2013representing,doi:10.1063/1.5027283}, as used in conventional interatomic potentials,
moments derived from the atomic distribution\cite{doi-10.1137-15M1054183}, 
the power spectrum, the bispectrum and the smooth overlap of atomic positions ({\sc SOAP}) kernel based on spherical harmonics\cite{kondor2008group,bartok2013representing}.
Simultaneously, group-theoretical methods have long been powerful tools for deriving invariants based on the symmetry of target systems, and some of the above invariants have been derived by group-theoretical methods.
Such group-theoretical invariants have been widely used not only for estimating MLIPs but also for characterizing and analyzing local structures (e.g., bond-orientational order parameter (BOP)\cite{PhysRevB.28.784}) and deriving the Landau free energy based on supergroup-subgroup relationship\cite{1987ltpt.book}, 
the potential energy surface for a molecule as a function of symmetry-adapted redundant coordinates\cite{collins1993implications}
and the model Hamiltonian of a crystalline system based on its space group\cite{PhysRevB.98.094105}.

In this study, we introduce high-order linearly independent rotational invariants up to the sixth order based on spherical harmonics into an MLIP framework, i.e., a linearized MLIP.
A set of rotational invariants is enumerated by the general process of reducing the Kronecker products of irreducible representations (Irreps) for the SO(3) group using a group-theoretical projector method.
As an application of high-order invariants, we demonstrate a linearized MLIP for elemental aluminum, formulated by a linear polynomial function of invariants.
Since the MLIP requires the highest accuracy among machine learning estimations of structure-property relationships, the enumeration of rotational invariants will be useful for constructing MLIPs with the desired accuracy.

\section{Potential energy models}
In this section, we introduce an atomic energy model with high-order polynomial invariants of the SO(3) group derived from the neighboring atomic density.
We first give a general formulation of the relationship between the atomic energy and the neighboring atomic density in Sec. \ref{sec-II1}.
Here, we consider the neighboring atomic density in a structure composed of a single element for simplicity.
The present formulation can be easily extended to multicomponent structures.
Then, we introduce a linear polynomial model for the atomic energy with polynomial invariants in Sec. \ref{sec-II2}.
In Sec. \ref{sec-II3}, a group-theoretical approach to construct linearly independent polynomial invariants of SO(3) is shown.
In Sec. \ref{sec-II4}, a practical form of radial functions used to expand the neighboring atomic density is demonstrated.
In the last two subsections, we show two other models for the atomic energy used for comparison with the model with high-order polynomial invariants.

\begin{figure}[tbp]
\includegraphics[clip,width=0.9\linewidth]{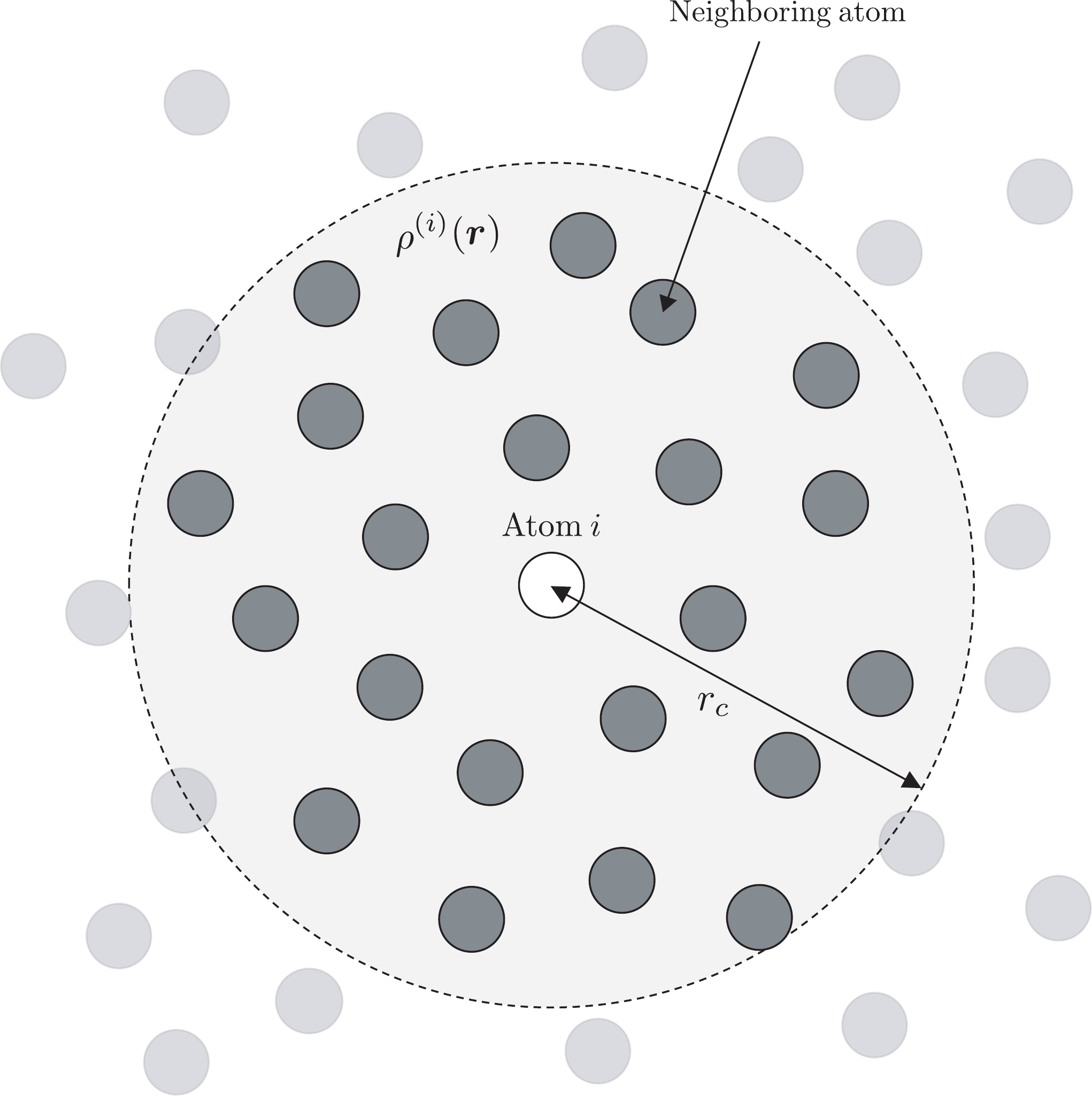}
\caption{
Schematic illustration of the neighboring atomic density around atom $i$ in a structure composed of a single element.
}
\label{mlip-gtinv-2018:atomic-distribution-schematic}
\end{figure}

\subsection{General formulation of interatomic potentials}
\label{sec-II1}
The total energy of a structure, $E$, may be divided into its atomic energies. 
This means that the total energy is expressed as
$E = \sum_i E^{(i)}$,
where $E^{(i)}$ denotes the contribution of atom $i$ to the total energy. 
Since the atomic energy depends only on its neighboring atoms, the relationship between the energy of atom $i$ and the neighboring atomic environment may be written in a functional form as
\begin{equation}
E^{(i)} = \mathcal{F} \left[ \rho^{(i)} \right], 
\end{equation}
where $\rho^{(i)}$ denotes the neighboring atomic density of atom $i$ illustrated schematically in Fig. \ref{mlip-gtinv-2018:atomic-distribution-schematic}.
In a structure composed of a single element, the neighboring atomic density is written as $\rho^{(i)} = \sum_{j \in {\rm neighbor}} \delta(\bm{r}-\bm{r}_j)$, where $\bm{r}_j$ denotes the position of neighboring atom $j$.
When we expand the neighboring atomic density in terms of a basis set $\{b_n\}$, the neighboring atomic density is rewritten as
\begin{equation}
\rho^{(i)} (\bm{r}) = \sum_n a_n^{(i)} b_n(\bm{r}),
\end{equation}
where $a_n^{(i)}$ are the order parameters. 
In this expansion, the set of order parameters identifies the neighboring atomic density.
Therefore, the atomic energy is a function of the set of order parameters.

Then, let us consider a situation where an arbitrary rotation is applied to the neighboring atomic density. 
We define rotation operator $\hat R$ acting on the basis functions as
\begin{eqnarray}
\hat R b_n^{(i)} (\bm{r}) &=& \sum_{n'} \bm{\Gamma}_{n'n} (\hat R) b_{n'}(\bm{r}) \nonumber \\
&=& b_n(\hat R ^{-1} \bm{r}),
\end{eqnarray}
where $\bm{\Gamma}(\hat R)$ denotes the matrix representation of rotation $\hat R$ for the basis set $\{b_n\}$.
This property holds for the atomic density expressed by a linear combination of the basis functions.
Therefore, the atomic density is transformed by rotation $\hat R$ to
\begin{equation}
\hat R \rho^{(i)} (\bm{r}) = \sum_{n,n'} a_{n}^{(i)} \bm{\Gamma}_{n'n} (\hat R) b_{n'}(\bm{r}).
\end{equation}
This transformation of the atomic density can be also viewed as the change of the order parameters from $a_n^{(i)}$ to $\sum_{n'} a_{n'}^{(i)} \bm{\Gamma}_{nn'}(\hat R)$. 

Although an arbitrary rotation generally changes the neighboring atomic density, it does not change the atomic energy.
Since all elements of the SO(3) group leave the atomic energy invariant, the atomic energy should be modeled by the invariants of the SO(3) group, $\{d_n^{(i)}\}$, derived from $\{a_n^{(i)}\}$. 
Finally, the atomic energy can be written as
\begin{equation}
E^{(i)} = F \left( d_1^{(i)}, d_2^{(i)}, \cdots \right).
\end{equation}
This formulation can describe both conventional interatomic potentials and MLIPs, and $\{d_n^{(i)}\}$ are called ``structural features'' in the context of the MLIP.

\subsection{Atomic energy model with high-order polynomial invariants}
\label{sec-II2}
Using the procedure introduced by Bart\'ok et al.\cite{bartok2013representing}, we expand the atomic density using a basis set corresponding to the Irreps of SO(3), i.e., spherical harmonics. 
In this case, a rotation of the basis set is represented by the Irreps of SO(3) known as the Wigner $D$-matrix.
A rotation of products of radial functions $\{f_n\}$ and spherical harmonics $\{Y_{lm}\}$ is also represented by the Irreps of SO(3).
When we expand the neighboring atomic density in terms of $\{f_n Y_{lm} \}$, the neighboring atomic density at a position $(r, \theta, \phi)$ in spherical coordinates centered at the position of atom $i$ is expressed as
\begin{equation}
\rho^{(i)} (r, \theta, \phi) = \sum_{nlm} a_{nlm}^{(i)} f_n(r) Y_{lm} (\theta, \phi),
\end{equation}
where $a_{nlm}^{(i)}$ are order parameters.
If one chooses an orthonormal set of radial functions, the order parameters $\{a_{nlm}^{(i)}\}$ can be calculated from the neighboring atomic density in a given structure using the relationship
\begin{equation}
a_{nlm}^{(i)} = \sum_{j \in \rm {neighbor}} f_n(r_{ij}) Y_{lm}^* (\theta_{ij}, \phi_{ij}),
\label{EquationOrderParameters}
\end{equation}
where $(r_{ij}, \theta_{ij}, \phi_{ij})$ denotes the spherical coordinates of neighboring atom $j$ centered at the position of atom $i$.


Then, we adopt linearly independent polynomial invariants of SO(3) generated from order parameters as structural features for the atomic energy.
A $p$th-order polynomial invariant for given $n$ and $\{l_1,l_2,\cdots,l_p\}$ is defined by a linear combination of products of $p$ order parameters, expressed as
\begin{equation}
d_{nl_1l_2\cdots l_p, (s)}^{(i)} = \sum_{m_1,m_2,\cdots, m_p} c^{l_1l_2\cdots l_p, (s)}_{m_1m_2\cdots m_p} a_{nl_1m_1}^{(i)} a_{nl_2m_2}^{(i)} \cdots a_{nl_pm_p}^{(i)}.
\end{equation}
As will be shown in Sec. \ref{sec-II3}, a coefficient set $\{c^{l_1l_2\cdots l_p, (s)}_{m_1m_2\cdots m_p}\}$ is constructed by solving the eigenvalue problem for a projector matrix, ensuring that the linear combination is invariant for arbitrary rotation.
Since there are no $l$-positive first-order invariants, all first-order invariants are identical to radial order parameters, $d_{n0}^{(i)} = a_{n00}^{(i)}$.
A second-order invariant is identified by a single value of $l$ because second-order linear combinations are invariant only when $l_1=l_2$, which is a general feature of product groups\cite{el-batanouny_wooten_2008,1987ltpt.book}.
In terms of fourth- and higher-order polynomial invariants, multiple invariants are linearly independent for most of the set $\{l_1,l_2,\cdots,l_p\}$, which are distinguished by index $s$ if necessary.

Finally, the atomic energy may be formulated as a function of the linearly independent polynomial invariants.
We employ an atomic energy model based on a simple linear polynomial form of the invariants, written as
\begin{widetext}
\begin{eqnarray}
E^{(i)} & = & 
w_0 + \sum_{n} w_{n0} d_{n0}^{(i)}
+ \sum_{n,l} w_{nll} d_{nll}^{(i)}
+ \sum_{n,\{l_1,l_2,l_3\}} w_{nl_1l_2l_3} d_{nl_1l_2l_3}^{(i)}
+ \sum_{n,\{l_1,l_2,l_3,l_4\},s} w_{nl_1l_2l_3l_4,s} d_{nl_1l_2l_3l_4,s}^{(i)}
+ \cdots ,
\label{Eqn-linear-polynomial}
\end{eqnarray}
\end{widetext}
where $w_0$, $w_{n0}$, $w_{nll}$, $w_{nl_1l_2l_3}$ and $w_{nl_1l_2l_3l_4,s}$ are regression coefficients.

Note that the second- and third-order invariants are equivalent to the angular Fourier series (AFS) and bispectrum reported in the literature, respectively\cite{kondor2008group,bartok2013representing}.
When excluding radial parts, the second order invariants are consistent with BOPs\cite{PhysRevB.28.784}.
If we restrict the invariants to third-order symmetrized ones excluding radial parts, they correspond to third-order BOPs.


\subsection{Group-theoretical projector operation method}
\label{sec-II3}
A set of polynomial invariants up to the sixth order is derived by the general process of reducing the Kronecker products of Irreps.
The Kronecker products of Irreps have been widely used for many purposes in physics and chemistry such as the formulation of angular momentum coupling, the derivation of selection rules and the formulation of the Landau free energy for phase transitions.
We employ the group-theoretical projection operator method\cite{el-batanouny_wooten_2008} to derive polynomial invariants of SO(3). 

Let us first consider the reduction of products of the same finite group $\mathcal{G}$.
In the projection operator method, the projector matrix of the Kronecker product of Irreps ${}^{(\mu)}\bm{\Gamma}$, ${}^{(\nu)}\bm{\Gamma}$, $\cdots$ of the same finite group $\mathcal{G}$ to Irrep $\sigma$ is defined by
\begin{eqnarray}
^{(\sigma)}\bm{P} &=& \sum_i {}^{(\sigma)}\bm{P}_{ii} \nonumber\\
&=& \frac{d_\sigma}{g} \sum_i \sum_{\hat R \in \mathcal{G}} {}^{(\sigma)}\Gamma_{ii}^{-1}(\hat R) {}^{(\mu)}\bm{\Gamma} (\hat R) \otimes {}^{(\nu)}\bm{\Gamma} (\hat R) \otimes \cdots,
\end{eqnarray}
where $g$ and $d_\sigma$ denote the order of group $\mathcal{G}$ and the dimension of Irrep $\sigma$, respectively.
Considering the reduction of the Kronecker product to the one-dimensional identity Irrep whose elements are all unity, the projection matrix is given by
\begin{equation}
{}^{(1)}\bm{P} = \frac{1}{g} \sum_{\hat R \in \mathcal{G}} {}^{(\mu)}\bm{\Gamma} (\hat R) \otimes {}^{(\nu)}\bm{\Gamma} (\hat R) \otimes \cdots.
\label{mlip-gtinv-2018:eqn-projector}
\end{equation}
By solving the eigenvalue problem for the projector matrix of each combination $\{\mu,\nu,\cdots\}$, expressed by  
\begin{equation}
{}^{(1)} \bm{P} \bm{u} = \bm{u},
\end{equation}
eigenvector $\bm{u}$ is obtained.
Each eigenvector corresponds to a set of coefficients identifying a polynomial invariant of group $\mathcal{G}$ in the form of a linear combination of order parameters.

SO(3) is an infinite or continuous group, which has an infinite number of elements.
In such a case, the average appearing in Eqn. (\ref{mlip-gtinv-2018:eqn-projector}) is replaced with an integral over continuous parameters\cite{Hamermesh:1123140}.
When a rotation is described by Euler angles $\{\alpha, \beta, \gamma\}$, the Wigner $D$-function $D^{(l)} (\alpha,\beta,\gamma)$ is an Irrep of SO(3).
Therefore, the element of the projector matrix for the decomposition of $p$th-order Kronecker products into the identity Irrep is written as
\begin{widetext}
\begin{equation}
{}^{(l=0)}P^{l_1l_2\cdots l_p}_{m_1m_2\cdots m_p,m_1'm_2'\cdots m_p'} =
\frac{1}{8\pi^2}\int_0^{2\pi} d\alpha \int_0^{\pi} d\beta \sin \beta \int_0^{2\pi} d\gamma D^{(l_1)}_{m_1m_1'} (\alpha,\beta,\gamma) D^{(l_2)}_{m_2m_2'} (\alpha,\beta,\gamma) \cdots D^{(l_p)}_{m_pm_p'} (\alpha,\beta,\gamma),
\end{equation}
\end{widetext}
where integrals are calculated only from Clebsch--Gordan coefficients for SO(3) or Wigner 3$j$ symbols as shown in Appendix \ref{sec-appendix-projector}.
The projector matrix element has a nonzero value only when $m_1+m_2+\cdots+m_p=0$ and $m_1'+m_2'+\cdots+m_p'=0$. 
Then, a solution of the eigenvalue problem for a given set $\{l_1,l_2,\cdots,l_p\}$ corresponds to a polynomial invariant of SO(3).

Note that some of the eigenvectors with nonzero eigenvalues correspond to invariants that are linearly dependent on invariants from the other eigenvectors. 
This originates from the fact that we consider Kronecker products of a single basis set.
For the same reason, some of the eigenvectors correspond to invariants that are constantly zero.
Therefore, they are removed from the set of invariants.
In addition, we use only polynomials that are invariant for an arbitrary improper rotation. 
This means that we use polynomial invariants for a set of $\{l_1,l_2,\cdots,l_p\}$ whose sum is even.

\subsection{Radial functions}
\label{sec-II4}

We employ Gaussian-type radial functions as $f_n(r)$ to define order parameters, where $f_n(r)$ is expressed as
\begin{equation}
f_{n}(r)=\exp\left[-\beta_n(r-r_n)^{2}\right] f_c(r),
\end{equation}
where $\beta_n$ and $r_n$ are given parameters.
Values of $\{\beta_n\}$ and $\{r_n\}$ are given by finite arithmetic progressions.
Considering the cutoff function $f_c$, we use the following cosine-based function:
\begin{eqnarray}
f_c(r) = \left\{
\begin{aligned}
& \frac{1}{2} \left[ \cos \left( \pi \frac{r}{r_c} \right) + 1\right] & (r \le r_c)\\
& 0 & (r > r_c)
\end{aligned}
\right .,
\end{eqnarray}
where $r_c$ denotes the cutoff radius.
Although Gaussian-type radial functions are not orthonormal, order parameters are approximately estimated using Eqn. (\ref{EquationOrderParameters}).

\subsection{Pair functional model}
\label{sec-II5}
For comparison with the atomic energy model with high-order polynomial invariants, we use the pair functional model introduced in Ref. \onlinecite{doi:10.1063/1.5027283}.
This model was reported to be capable of predicting the energy within a root mean square (RMS) error of 2.7 meV/atom for datasets derived from simple structure generators, which were averaged over 31 elemental metals including transition metals\cite{doi:10.1063/1.5027283}.
We use the second-order approximation of the pair functional model described as
\begin{equation}
E^{(i)} = w_0 + \sum_n w_{n0} d_{n0}^{(i)} + \sum_{n,n'} w_{n0,n'0} d_{n0}^{(i)}d_{n'0}^{(i)}.
\end{equation}

Because MLIPs are generally regarded as extensions of conventional interatomic potentials, a classification rule of conventional interatomic potentials based on the type of structural features\cite{carlsson1990beyond} is applicable to MLIPs.
This model is classified into a pair functional potential; hence, we call this model a pair functional MLIP hereafter.

\subsection{Cluster functional model}
\label{sec-II6}
A cluster functional model with AFS structural features is also introduced for comparison. 
The AFS is given by
\begin{equation}
d_{nl}^{(i)} = \sum_{j,k \in {\rm neighbor}} f_n(r_{ij})f_n(r_{ik}) \cos (l \theta_{ijk}),
\end{equation}
where $\theta_{ijk}$ denotes the bond angle among atom $i$ and its neighboring two atoms.
When AFSs are used as structural features in the linear polynomial model of Eqn. (\ref{Eqn-linear-polynomial}), they are derived to be equivalent to second-order polynomial invariants using the addition theorem of spherical harmonics\cite{bartok2013representing}.

Here, we employ a second-order polynomial approximation with AFS structural features written as
\begin{equation}
E^{(i)} = w_0 + \displaystyle\sum_{n,l} w_{nl} d_{nl}^{(i)} + \sum_{n,l,n',l'} w_{nl,n'l'} d_{nl}^{(i)}d_{n'l'}^{(i)},
\end{equation}
where Gaussian-type radial functions are adopted.
This model was also applied to the 31 elemental metals using the dataset mentioned in the previous subsection\cite{doi:10.1063/1.5027283}.
The RMS error averaged over the 31 elemental metals was reported to be 0.5 meV/atom.
We call this model a cluster functional MLIP because it is classified as a cluster functional potential.

\section{Estimation of potential energy models}
\subsection{Datasets}
\label{Sec-Datasets}

Training and test datasets are constructed from DFT calculations.
The test dataset is used to estimate the predictive power for structures that are not included in the training dataset.
To generate a wide range of structures for the training and test datasets, we adopt prototype structures reported in the Inorganic Crystal Structure Database (ICSD)\cite{bergerhoff1987crystal} as structure generators.
We restrict all ICSD entries to those with the ANX formula of ``N'' and eliminate duplicate entries that have the same ICSD ``structure type". 
In other words, we employ only unique ICSD prototype structures composed of single elements with zero oxidation state.
The total number of structure generators is 86 and a list of structure generators is shown in Appendix \ref{appendix-structure-generator}.

First, the atomic positions and lattice constants of the structure generators are fully optimized by DFT calculation to obtain their equilibrium structures.
Then, a candidate structure used in each of the datasets is constructed by random lattice expansion, random lattice distortion and random atomic displacements into a supercell of each of the structure generators.
For a given parameter $\varepsilon$ controlling the degree of lattice expansion, lattice distortion and atomic displacements, the lattice vectors of the candidate structure are expressed by
\begin{equation}
\bm{A'} = \bm{A} + \varepsilon\bm{R},
\end{equation}
where $\bm{A}$ and $\bm{A'}$ denote the matrix representations of the lattice vectors of the supercell of the structure generator and the candidate structure, respectively.
The $(3\times3)$ matrix $\bm{R}$ is composed of uniform random numbers ranging from $-1$ to 1.
The displacement of an atom is described by the change of its fractional coordinates as 
\begin{equation}
\bm{f'} = \bm{f} + \varepsilon \bm{A'}^{-1} \bm{\eta},
\end{equation}
where $\bm{f}$ and $\bm{f'}$ denote the fractional coordinates of the atom in the supercell of the structure generator and the candidate structure, respectively.
The three-dimensional vector $\bm{\eta}$ consists of uniform random numbers ranging from $-1$ to 1.

We generate a wide range of candidate structures using multiple values of $\varepsilon$.
When we generate $N_{\rm st}$ structures from a structure generator, the value of $\varepsilon$ for the $N$th structure, $\varepsilon_N$, is given by the finite arithmetic progression of length $N_{\rm st}$ as $\varepsilon_N = 0.5N / N_{\rm st}$ \AA.
Therefore, it is worth emphasizing that the given displacements are much larger than the small displacements required to compute harmonic phonon force constants.
Applying this procedure to all the structure generators, a total of 430,000 candidate structures are generated.

Among the candidates, we sample 10,000 structures so that the variance of the predicted energy becomes small in a similar manner to structure selection procedures in the cluster expansion for alloy systems\cite{CV2,sampling:seko,PhysRevB.82.184107}. 
We then split them into 9000 and 1000 structures for the training and test datasets, respectively.
For the total of 10,000 structures, DFT calculations were performed using the plane-wave-basis projector augmented wave (PAW) method\cite{PAW1} within the Perdew--Burke--Ernzerhof exchange-correlation functional\cite{GGA:PBE96} as implemented in the \textsc{VASP} code\cite{VASP1,VASP2,PAW2}.
The cutoff energy was set to 400 eV.
The total energies converged to less than 10$^{-3}$ meV/supercell.
The atomic positions and lattice constants were optimized for the structure generators until the residual forces were less than 10$^{-2}$ eV/\AA.

\subsection{Regression}
\label{sec-regression}
Regarding the training data, the total energy, the forces acting on atoms and the stress tensor computed by DFT calculations are available since they are all expressed by linear equations with the same regression coefficients.
When we estimate the regression coefficients from all of the total energies, forces and stress tensors, the predictor matrix $\bm{X}$ is divided into three submatrices $\bm{X}_{\rm energy}$, $\bm{X}_{\rm force}$ and $\bm{X}_{\rm stress}$, which contain the structural features for the total energies, the forces acting on atoms and the stress tensors of structures in the training dataset, respectively.
The structural features for the forces are derived in Appendix \ref{appendix-force}, and the structural features for the stress tensors can be easily derived in a similar manner to the derivation of the structural features for the forces.
The observation vector also has three components of the total energy $\bm{y}_{\rm energy}$, forces $\bm{y}_{\rm force}$ and stress tensor $\bm{y}_{\rm stress}$ of structures in the training dataset, obtained by DFT calculations.
The predictor matrix and observation vector are simply written in a submatrix form as
\begin{equation}
\bm{X} =
\begin{bmatrix}
\bm{X}_{\rm energy} \\
\bm{X}_{\rm force} \\
\bm{X}_{\rm stress} \\
\end{bmatrix}
,\qquad \bm{y} =
\begin{bmatrix}
\bm{y}_{\rm energy} \\
\bm{y}_{\rm force} \\
\bm{y}_{\rm stress} \\
\end{bmatrix}.
\end{equation}
Finally, the total number of training data reached 1,377,769.

The regression coefficients of a model for the atomic energy, which comprise the coefficient vector $\bm{w}$, are estimated by linear ridge regression. 
The optimal ridge coefficients minimize the penalized residual sum of squares expressed as
\begin{equation}
L(\bm{w}) = ||\bm{X}\bm{w} - \bm{y}||^2_2 + \lambda ||\bm{w}||^2_2,
\end{equation}
where $\bm{X}$ and $\bm{y}$ denote the predictor matrix and observation vector, respectively.
The regularization parameter $\lambda$ controls the magnitude of the penalty.
It is also beneficial to use sparse linear regressions such as the least absolute shrinkage and selection operator (Lasso) to decrease the cost of computing the energy and forces, while we adopt the linear ridge regression to estimate the regression coefficients rapidly and stably in this study.

\section{Results and discussion}
\subsection{Number of invariants}
The number of polynomial invariants that occur in the decomposition of the Kronecker product of the Irreps of the SO(3) group is obtained only from their characters.
The number of $p$th-order polynomial invariants for the set $\{l_1,l_2,\cdots, l_p\}$ is calculated using the following equation:
\begin{widetext}
\begin{equation}
n_{l_1,l_2,\cdots,l_p} =
\frac{1}{8\pi^2}\int_0^{2\pi} d\alpha \int_0^{\pi} d\beta \sin \beta \int_0^{2\pi} d\gamma \chi^{(l_1)} (\alpha,\beta,\gamma) \chi^{(l_2)} (\alpha,\beta,\gamma) \cdots \chi^{(l_p)} (\alpha,\beta,\gamma),
\end{equation}
\end{widetext}
where $\chi^{(l)} (\alpha,\beta,\gamma)$ denotes the character of Irrep $l$ for a rotation described by Euler angles $\{\alpha,\beta,\gamma \}$.
In a practical enumeration of invariants, the explicit forms of characters described by its rotation axis $\bm{\phi}$ and rotation angle $\omega$ are more convenient than those described by Euler angles.
When specifying the rotation in this manner, the character of Irrep $l$ for rotation $\hat R$ is simply expressed as
\begin{equation}
\chi^{(l)} (\hat R) = \chi^{(l)}(\omega) = \frac{\sin \left[ (2l+1) \omega / 2 \right]}{\sin \left[\omega / 2 \right]}.
\end{equation}
Therefore, the number of $p$th-order polynomial invariants is computed as
\begin{widetext}
\begin{equation}
n_{l_1,l_2,\cdots,l_p} =
\frac{1}{4\pi^2}
\int_0^{2\pi} d\omega \sin^2 \frac{\omega}{2} \int_0^{\pi} d\theta \sin \theta \int_0^{2\pi} d\phi \chi^{(l_1)} (\omega) \chi^{(l_2)} (\omega) \cdots \chi^{(l_p)} (\omega),
\end{equation}
\end{widetext}
where $\theta$ and $\phi$ identify the rotation axes.

\begin{table}[tbp]
\begin{ruledtabular}
\caption{
Cumulative number of SO(3) invariants for a given value of maximum $l$.
}
\label{mlip-gtinv-2018:number-of-invariants}
\begin{tabular}{r|rrrrr}
         & \multicolumn{5}{c}{Order of polynomial invariant} \\
 $l_{\rm max}$ & 2 & 3 & 4 & 5 & 6 \\
\hline
 0   &1	    &1      & 1      &  1       &1     \\
 1   &2	    &3      & 6      &  12      &27     \\
 2   &3	    &7      & 23     &  79      &295     \\
 3   &4	    &13     & 65     &  336     &1841     \\
 4   &5	    &22     & 153    &  1102    &8222      \\
 5   &6	    &34     & 317    &  3019    &29274     \\
 6   &7	    &50     & 598    &  7257    &88402     \\
 7   &8	    &70     & 1049   &  15778   &235439    \\
 8   &9	    &95     & 1738   &  31692   &567795    \\
 9   &10	&125    & 2748   &  59688   &1263347   \\
 10  &11	&161 	& 4180   &  106580  &2629421   \\
 11  &12	&203 	& 6154   &  181947  &5173123   \\
 12  &13	&252 	& 8811   &  298910  &9699354   \\
 13  &14	&308 	& 12314  &  475021  &17443960  \\
 14  &15	&372 	& 16851  &  733313  &30250729  \\
 15  &16	&444 	& 22635  &  1103478 &50802227  \\
 16  &17	&525 	& 29907  &	1623230 &82915955  \\
 17  &18	&615 	& 38937  &	2339815 &131918757 \\
 18  &19	&715 	& 50026  &	3311727 &205114070 \\
 19  &20	&825 	& 63507  &	4610589 &312358280 \\
 20  &21	&946 	& 79748  &	6323265 &466764285 \\
\end{tabular}
\end{ruledtabular}
\end{table}

\begin{table}[tbp]
\begin{ruledtabular}
\caption{
Cumulative number of nonzero linearly independent SO(3) invariants derived from products of order parameters of Eqn. (\ref{EquationOrderParameters}).
}
\label{mlip-gtinv-2018:number-of-invariants2}
\begin{tabular}{r|rrrrr}
         & \multicolumn{5}{c}{Order of polynomial invariant} \\
 $l_{\rm max}$ & 2 & 3 & 4 & 5 & 6 \\
\hline
 0   &1   &1    &1     &1    & 1  \\
 1   &2   &2    &3     &3    & 4  \\
 2   &3   &5    &9     &13   & 23 \\
 3   &4   &8    &26    &53   & 146\\
 4   &5   &15   &64    &218  & $-$\\
 5   &6   &22   &136   &681  & $-$\\
 6   &7   &35   &273   &1919 & $-$\\
 7   &8   &48   &500   &$-$  & $-$\\
 8   &9   &69   &864   &$-$  & $-$\\
 9   &10  &90   &1423  &$-$  & $-$\\
 10  &11  &121  &2246  &$-$  & $-$\\
 11  &12  &152  &$-$  &$-$  & $-$\\
 12  &13  &195  &$-$  &$-$  & $-$\\
 13  &14  &238  &$-$  &$-$  & $-$\\
 14  &15  &295  &$-$  &$-$  & $-$\\
 15  &16  &352  &$-$  &$-$  & $-$\\
 16  &17  &425  &$-$  &$-$  & $-$\\
 17  &18  &498  &$-$  &$-$  & $-$\\
 18  &19  &589  &$-$  &$-$  & $-$\\
 19  &20  &680  &$-$  &$-$  & $-$\\
 20  &21  &791  &$-$  &$-$  & $-$\\
\end{tabular}
\end{ruledtabular}
\end{table}

Table \ref{mlip-gtinv-2018:number-of-invariants} shows the number of $p$th-order invariants satisfying $l_1 \leq l_{\rm{max}}$, $l_2 \leq l_{\rm{max}}$, $\cdots$, $l_p \leq l_{\rm{max}}$ for a given $l_{\rm{max}}$.
The integer sequences for second- and third-order invariants may correspond to the On-Line Encyclopedia of Integer Sequences (OEISs) A000027 and A002623, respectively\cite{oeis}.
Nonetheless, as described above, these numbers include linearly dependent and constantly zero invariants derived from a single basis set.
Therefore, the number of available invariants shown in Table \ref{mlip-gtinv-2018:number-of-invariants2} is obtained by solving the eigenvalue problems for the projector matrix and removing such invariants.

\begin{table}[tbp]
\begin{ruledtabular}
\caption{
Cumulative number of symmetrized invariants for a given value of maximum $l$, $l_{\rm max}$.
}
\label{mlip-gtinv-2018:number-of-sym-invariants}
\begin{tabular}{r|rrrrr}
& \multicolumn{5}{c}{Order of polynomial invariant} \\
$l_{\rm max}$ & 2 & 3 & 4 & 5 & 6 \\
\hline
 0   &1   &1   &1   &1   &1   \\
 1   &2   &1   &2   &1   &2   \\
 2   &3   &2   &3   &2   &4   \\
 3   &4   &2   &5   &2   &7   \\
 4   &5   &3   &7   &4   &11  \\
 5   &6   &3   &9   &4   &17  \\
 6   &7   &4   &12  &7   &25  \\
 7   &8   &4   &15  &7   &35  \\
 8   &9   &5   &18  &11  &48  \\
 9   &10  &5   &22  &12  &64  \\
 10  &11  &6   &26  &17  &84  \\
 11  &12  &6   &30  &18  &108 \\
 12  &13  &7   &35  &25  &137 \\
 13  &14  &7   &40  &27  &171 \\
 14  &15  &8   &45  &35  &211 \\
 15  &16  &8   &51  &38  &258 \\
 16  &17  &9   &57  &48  &312 \\
 17  &18  &9   &63  &52  &374 \\
 18  &19  &10  &70  &64  &445 \\
 19  &20  &10  &77  &69  &525 \\
 20  &21  &11  &84  &83  &616 
\end{tabular}
\end{ruledtabular}
\end{table}

Although we have considered invariants for all possible combinations of $l$ thus far, we are also allowed to restrict them to symmetrized invariants, which are derived by reducing the $p$th power of Irrep $l$\cite{el-batanouny_wooten_2008}.
The symmetrized power of an Irrep has played an important role in many subjects such as the construction of symmetrized tensors and the Landau theory of phase transitions.
The number of $p$th-order symmetrized invariants is expressed as
\begin{equation}
n_{l,p}^{\rm sym} =
\frac{1}{4\pi^2}
\int_0^{2\pi} d\omega \sin^2 \frac{\omega}{2} \int_0^{\pi} d\theta \sin \theta \int_0^{2\pi} d\phi \chi [D^{(l)}(\omega)^p],
\end{equation}
where $\chi[D^{(l)}(\omega)^p]$ denotes the character of the symmetrized $p$th power of Irrep $l$.
Appendix \ref{appendix-sym-character} presents the relationship between the character of an Irrep and that of the symmetrized $p$th power of the Irrep.

Table \ref{mlip-gtinv-2018:number-of-sym-invariants} shows the cumulative number of symmetrized invariants up to a given maximum $l$.
The integer sequences for second- and third-order invariants correspond to OEISs A000027 and A008619, respectively\cite{oeis}.
The increments of the integer sequences for second-, third-, fourth-, fifth- and sixth-order invariants also correspond to the first 20 elements of OEIS A000012, A059841, A008620, A008743 and A008669, respectively\cite{oeis}. 
The increment from $l_{\rm max}-1$ to $l_{\rm max}$ for order $p$ is consistent with the number of symmetrized invariants derived from the $p$th power of Irrep $l_{\rm max}$.

\subsection{Application to elemental aluminum}
We first demonstrate a procedure to construct an optimal MLIP with polynomial invariants for elemental Al.
An MLIP is identified by a given maximum value of order $p$, $p_{\rm max}$, and a given set of maximum values of $l$ for order $p$, $\{l_{\rm max}^{(2)}, l_{\rm max}^{(3)}, \cdots, l_{\rm max}^{(p_{\rm max})}\}$.
This means that all polynomial invariants satisfying $p \leq p_{\rm max}$ and $l_1 \leq l_{\rm{max}}^{(p)}$, $l_2 \leq l_{\rm{max}}^{(p)}$, $\cdots$, $l_p \leq l_{\rm{max}}^{(p)}$ are included in the set of structural features for a given $p_{\rm max}$ and $\{l_{\rm max}^{(2)}, l_{\rm max}^{(3)}, \cdots, l_{\rm max}^{(p_{\rm max})}\}$.
We hereafter describe an MLIP as ($l_{{\rm max}}^{(2)}$-$l_{{\rm max}}^{(3)}$-$\cdots$-$l_{{\rm max}}^{(p_{\rm max})}$).
To find an optimal MLIP, we consider $l_{\rm max}^{(p)}$ up to nine for the second order, seven for the third order, three for the fourth order and two for both the fifth and sixth orders.
For symmetrized invariants, $l_{\rm max}^{(p)}$ up to nine for the second order, eight for the third order, eight for the fourth order, four for the fifth order and three for the sixth order are taken into account.
In addition, the number of radial functions, the parameters in the radial functions and the cutoff radius are optimized by trial and error.
As a result, we use 22 sets of parameters $\beta_n$ and $r_n$ corresponding to all combinations of finite arithmetic progressions of $\beta_n = \{0.25, 0.5 \}$ and $r_n = \{0, 1, \cdots, 10\}$.
The cutoff radius is set to 10 \AA.

\begin{figure}[tbp]
\includegraphics[clip,width=\linewidth]{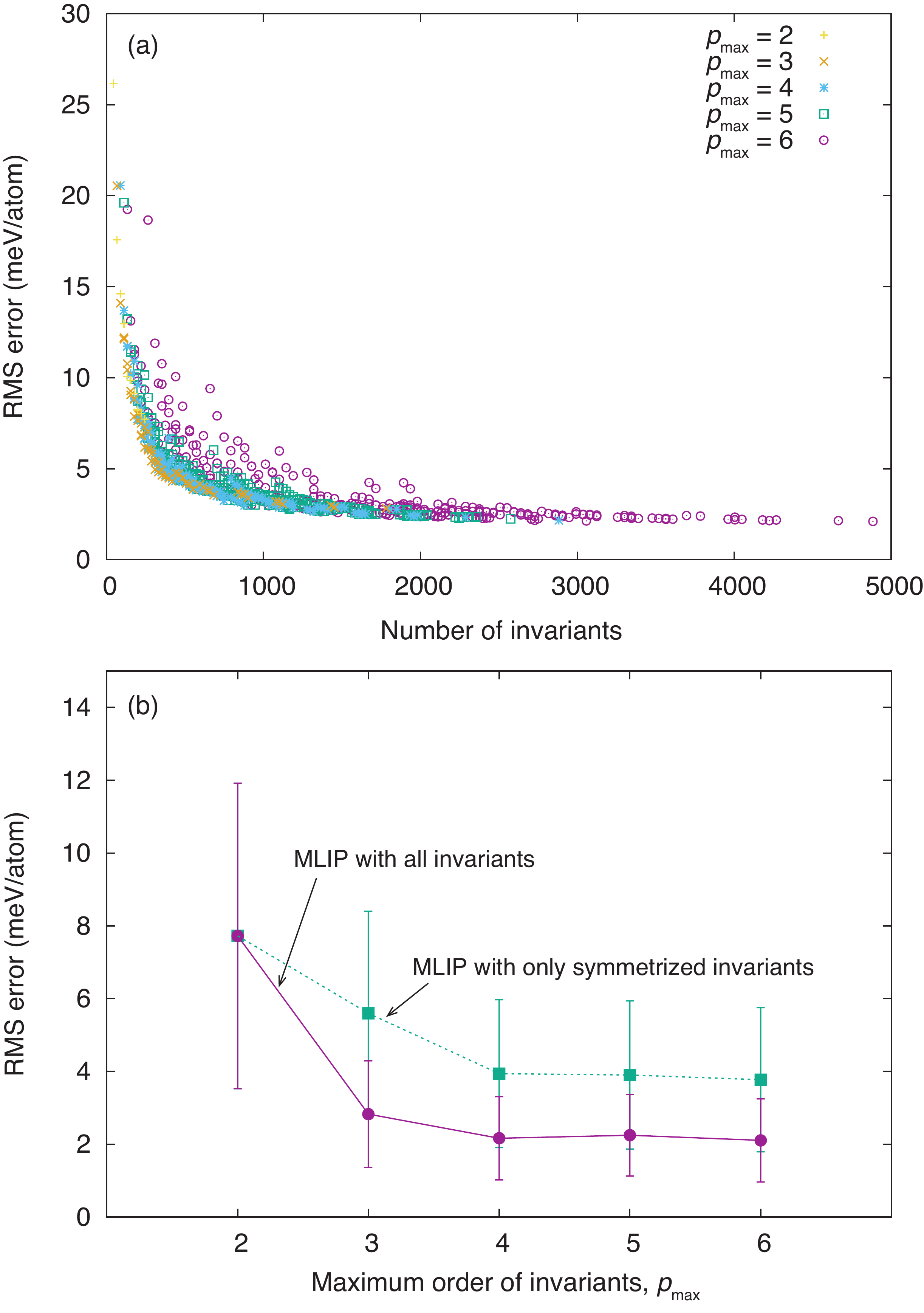}
\caption{
(a) Dependence of the RMS error on the number of invariants for Al.
(b) Dependence of the RMS error on the maximum order of invariants.
Error bars indicate standard deviations of RMS errors for structure groups.
The RMS errors of MLIPs composed of all invariants and those of MLIPs composed of only symmetrized invariants are shown by purple closed circles and green closed squares, respectively.
}
\label{mlip-gtinv-2018:nfeatures-rmse-Al}
\end{figure}

Figure \ref{mlip-gtinv-2018:nfeatures-rmse-Al} (a) shows the dependence of the RMS error for the test data on the number of invariants.
Each data point corresponds to the RMS error of an MLIP $(l_{\rm max}^{(2)}, l_{\rm max}^{(3)}, \cdots, l_{\rm max}^{(p_{\rm max})})$.
As can be seen in Fig. \ref{mlip-gtinv-2018:nfeatures-rmse-Al} (a), the RMS error decreases as the number of invariants increases.
The MLIP with the lowest RMS error of 2.27 meV/atom is composed of 2399 invariants (9-7-3-2-2).
The fitting error for the training data of 2.23 meV/atom is almost the same as the RMS error for the test data.
%

On the other hand, the pair functional MLIP has a large RMS error of 19.88 meV/atom although it was reported to have an RMS error of 0.89 meV/atom for the test data constructed from simple structure generators such as fcc-, bcc- and hcp-type structures\cite{doi:10.1063/1.5027283}.
Clearly, the difference in RMS error originates from the different structures used to estimate the prediction error, because the present test data is obtained from a wide range of structure generators as shown in Sec. \ref{Sec-Datasets}.
Therefore, the pair functional MLIP provides an accurate description of the atomic interactions in simple structures, whereas it has no power to describe the atomic interactions for a wide range of structures.
This limitation of the transferability has also long been recognized in conventional pair functional interatomic potentials such as embedded atom method (EAM) potentials, which are regarded as reductions of the pair functional MLIP.
Although the use of many pairwise structure features in the pair functional MLIP improves the descriptive power of the pair functional model for simple structures, this result indicates that systematically increasing the number of pairwise structural features is not a useful way of increasing the transferability to a wide range of structures.

Figure \ref{mlip-gtinv-2018:nfeatures-rmse-Al} (b) shows the dependence of the RMS error on the maximum order of the invariants. 
Only the MLIP with the lowest RMS error among all MLIPs identified by a value of $p_{\rm max}$ is shown in Fig. \ref{mlip-gtinv-2018:nfeatures-rmse-Al} (b).
The best MLIP composed of only second-order invariants, which is equivalent to an MLIP with AFS structural features, has an RMS error of 7.72 meV/atom.
Using both second- and third-order invariants, the RMS error significantly decreases to 2.83 meV/atom.
Upon adding higher-order invariants, the RMS error gradually decreases and MLIP (9-7-3-2-2) exhibits the lowest RMS error of 2.10 meV/atom.
Even when restricting the invariants to symmetrized ones, the RMS error decreases with increasing maximum order of the invariants.
MLIP (9-8-8-4-3) composed of only symmetrized invariants has an RMS error of 3.77 meV/atom, which is the lowest among the MLIPs with only symmetrized invariants.

However, the RMS error almost converges at $p_{\rm max}=4$. 
To further increase the accuracy of the MLIP, both a large number of additional high-order invariants and a large number of additional training data may be required, because the RMS error decreases very slowly as the number of invariants increases as shown in Fig. \ref{mlip-gtinv-2018:nfeatures-rmse-Al} (a).
In addition, the potential performance of higher-order invariants may be evaluated more accurately simply by significantly increasing the number of training data and high-order invariants.

\begin{figure*}[tbp]
\includegraphics[clip,width=\linewidth]{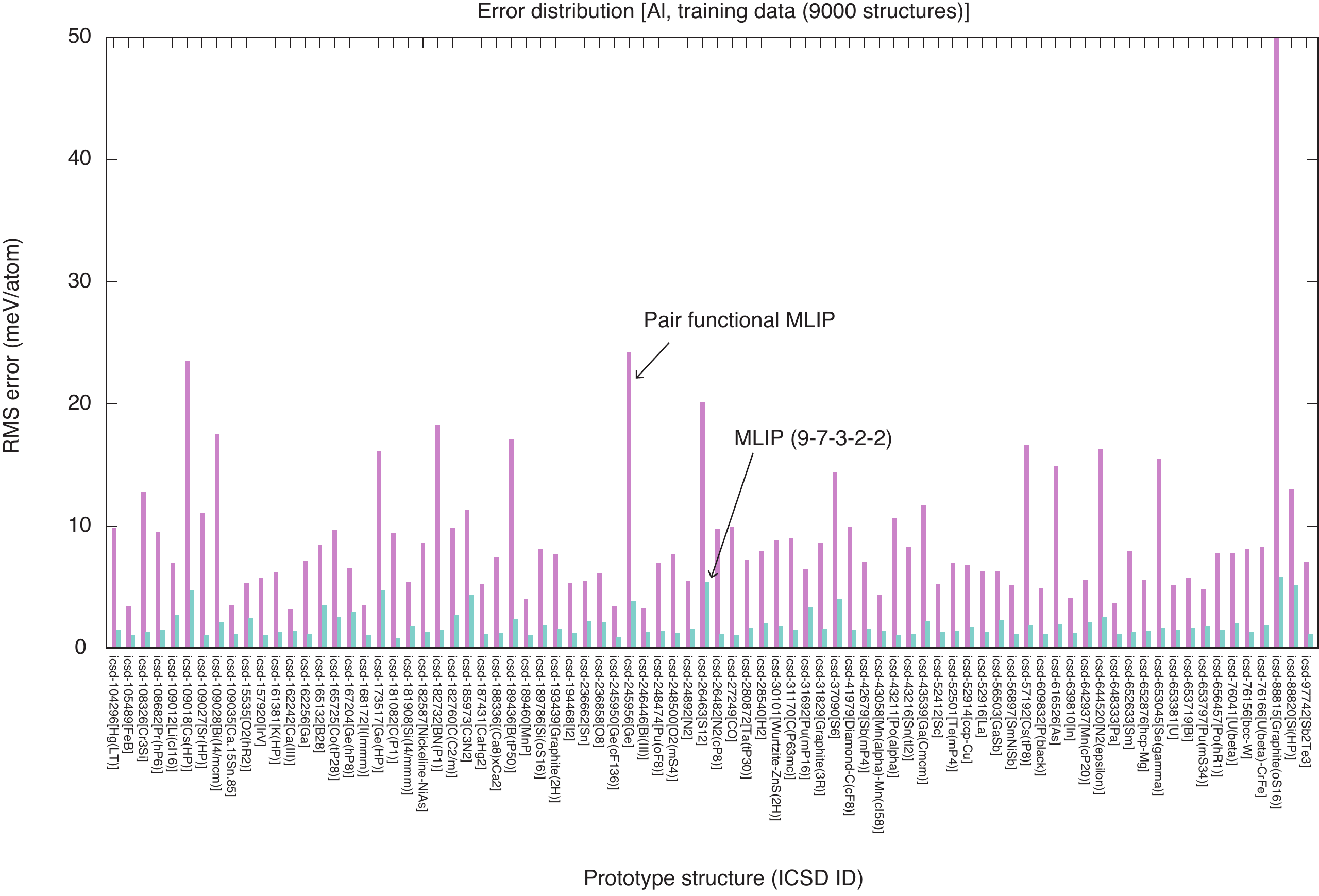}
\caption{
Dependence of fitting error on the structure group for Al for the pair functional MLIP and MLIP (9-7-3-2-2) shown by purple and green bars, respectively.
The prediction error for test data also shows a similar behavior to the fitting error.
The structure group is identified by ICSD-ID and the structure type listed in the longview of the ICSD.
}
\label{mlip-gtinv-2018:error_distribution_prototype-Al}
\end{figure*}

We measure in more detail the accuracy for a structure group, which is defined as a set of structures derived from a structure generator.
Figure \ref{mlip-gtinv-2018:error_distribution_prototype-Al} shows the dependence of the RMS error on the structure group of MLIP (9-7-3-2-2) in comparison with that of the pair functional MLIP.
The RMS error for a structure group is averaged over the structures in the group.
The pair functional MLIP exhibits large errors for some structure groups, such as high-pressure and covalent structures, and shows relatively small errors for the other structure groups.
On the other hand, the use of high-order polynomial invariants significantly improves the accuracy for the whole range of structure groups.
Figure \ref{mlip-gtinv-2018:nfeatures-rmse-Al} (b) also shows the standard deviation of the errors for the structure groups.
Figure \ref{mlip-gtinv-2018:nfeatures-rmse-Al} (b) indicates that high-order polynomial invariants decrease not only the error averaged over all structures but also the structure group dependence of the error.
In addition to the energy, the other physical properties depending on the structure group should be predicted with a higher accuracy by MLIP (9-7-3-2-2).

\begin{figure}[tbp]
\includegraphics[clip,width=\linewidth]{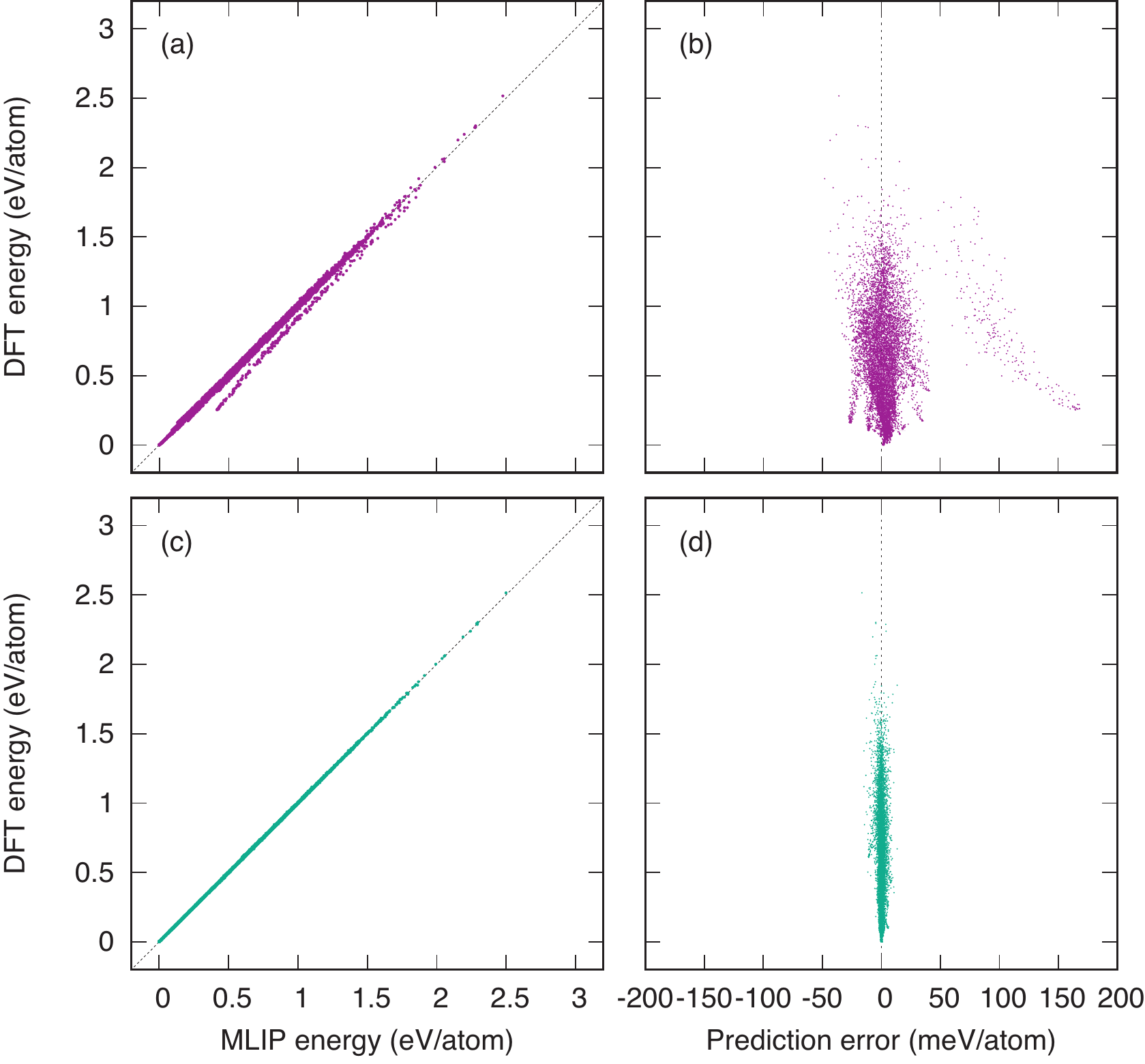}
\caption{
(a) Comparison of energies predicted by the pair functional MLIP with those predicted by DFT calculation.
The diagonal line indicates that the MLIP energy is equal to the DFT energy.
The MLIP and DFT energies of structures in the whole dataset are measured from those of equilibrium fcc structures.
(b) Distribution of the prediction error of the pair functional MLIP for structures in the whole dataset, along with their DFT energy.
(c) Comparison of energies predicted by MLIP (9-7-3-2-2) with those predicted by the DFT calculation.
(d) Distribution of the prediction error of MLIP (9-7-3-2-2) for structures in the whole dataset, along with their DFT energy.
}
\label{mlip-gtinv-2018:error_distribution-Al}
\end{figure}

Figure \ref{mlip-gtinv-2018:error_distribution-Al} shows a comparison of the energy computed by DFT calculation with those predicted by (a) the pair functional MLIP and (c) MLIP (9-7-3-2-2) for the whole dataset combining the training and test datasets.
The energy is measured from the DFT energy of the fcc equilibrium structure.
In the comparison of the pair functional MLIP with the DFT calculation, a series of structures shows a systematic deviation from the diagonal line indicating that the MLIP energy is equal to the DFT energy, although most of the structures are distributed around the diagonal line.
On the other hand, no such systematic deviation is seen in the comparison of MLIP (9-7-3-2-2) with the DFT calculation. 
The DFT and MLIP energies are almost the same for all structures in the energy scale of Fig. \ref{mlip-gtinv-2018:error_distribution-Al} (c).

Figure \ref{mlip-gtinv-2018:error_distribution-Al} also shows the energy dependence of the error for (b) the pair functional MLIP and (d) MLIP (9-7-3-2-2). 
It is clearly found that the pair functional MLIP has a wider error distribution than MLIP (9-7-3-2-2).
Although a clear energy dependence of the error is not observed in Figs. \ref{mlip-gtinv-2018:error_distribution-Al} (b) and (d), the error of the structures with a low DFT energy is much lower than those of the other structures.
Moreover, there are several other systematic deviations in the error distribution of the pair functional MLIP in addition to a clear systematic deviation that can also be recognized in Fig. \ref{mlip-gtinv-2018:error_distribution-Al} (a).

\begin{table}[tbp]
\begin{ruledtabular}
\caption{
Transferability of pair functional MLIP, cluster functional MLIP with AFS and MLIP (9-7-3-2-2) in elemental Al.
}
\label{mlip-gtinv-2018:transferability}
\begin{tabular}{lcccc}
& Pair func. & Cluster func. & MLIP        & DFT \\
&            & (AFS)      & (9-7-3-2-2) & \\
\hline
\multicolumn{5}{l}{RMS error in phonon frequency (THz)} \\
FCC & 0.118 & 0.177 & 0.052 & $-$ \\
BCC & 0.282 & 0.423 & 0.289 & $-$ \\
HCP & 0.185 & 0.921 & 0.053 & $-$ \\
\hline
\multicolumn{5}{l}{Vacancy formation energy (eV)} \\
FCC & 0.727 & 0.669 & 0.660 & 0.614 \\
HCP & 0.652 & 0.791 & 0.681 & 0.585 \\
\hline
\multicolumn{5}{l}{Grain boundary energy (mJ/m$^2$), STGB(110)} \\
\{113\} (50.479$^\circ$) & 187.5 & 179.4 & 187.4 & 158.4 \\
\{112\} (70.529$^\circ$) & 376.6 & 354.3 & 379.8 & 348.3 \\
\{111\} (109.471$^\circ$) & 88.8  & 78.1  & 57.5  & 39.2  \\
\{221\} (141.058$^\circ$) & 397.0 & 409.1 & 421.2 & 386.0 \\
\end{tabular}
\end{ruledtabular}
\end{table}

Undoubtedly, a useful MLIP requires not only a low prediction error but also predictive power for properties related to the energies of structures that are not included in the datasets, i.e., transferability.
Here, we evaluate the prediction error for the phonon frequencies, vacancy formation energy and grain boundary energy to estimate the transferability of MLIPs.
Table \ref{mlip-gtinv-2018:transferability} shows the RMS error for the phonon frequencies, vacancy formation energy and symmetric tilt grain boundary (STGB) energy.
The RMS error for the phonon frequency is evaluated as
\begin{equation}
({\rm RMS_{phonon}})^2 = \frac{1}{N_{\rm kpoints}} \sum_{k,n} \left( \omega_{k,n}^{\rm MLIP} - \omega_{k,n}^{\rm DFT} \right)^2,
\end{equation}
where $\omega_{k,n}^{\rm MLIP}$ and $\omega_{k,n}^{\rm DFT}$ denote the phonon frequencies of band $n$ at $k$-point $k$ predicted by the MLIP and DFT calculation, respectively.
Table \ref{mlip-gtinv-2018:transferability} indicates that the phonon frequencies predicted by MLIP (9-7-3-2-2) are very close to those predicted by DFT calculation.
The pair functional MLIP also has good predictive power for the phonon properties of fcc, bcc and hcp structures, although it has large errors ranging from 5 to 8 meV/atom for fcc, bcc and hcp structure groups as shown in Fig. \ref{mlip-gtinv-2018:error_distribution_prototype-Al}.
The high predictive power for the phonon properties of the pair functional MLIP is ascribed to the fact that structures with a low DFT energy have very small prediction errors in the pair functional MLIP as shown in Fig. \ref{mlip-gtinv-2018:error_distribution-Al} (b).

We next compare the vacancy formation energies for fcc and hcp structures predicted by MLIPs with those predicted by DFT calculation.
Supercells are formed by the $3\times3\times3$ expansions of the conventional unit cells of the fcc and hcp structures.
As shown in Table \ref{mlip-gtinv-2018:transferability}, all MLIPs show similar predictive powers for the vacancy formation energy.
We also examine the prediction error for grain boundary models not included in the training and test datasets.
Here, we introduce STGB models with a $\left<110\right>$ tilt direction. 
They have $\left\{113\right\}$, $\left\{112\right\}$, $\left\{111\right\}$ and $\left\{221\right\}$ grain boundary planes and misorientation angles of 50.479, 70.529, 109.471 and 141.058$^\circ$, respectively.
They are respectively composed of 880, 284, 384 and 512 atoms including two boundary planes.
Starting from their initial structures taken from Ref. \onlinecite{tschopp2015symmetric}, we optimize the STGB models by using MLIPs and by DFT calculation.
Table \ref{mlip-gtinv-2018:transferability} shows the grain boundary energies of the STGB models predicted by DFT calculation and MLIPs.
As can be seen in Table \ref{mlip-gtinv-2018:transferability}, all MLIPs predict the grain boundary energy very accurately.
These results are associated with the above discussion on the high predictive power for the phonon properties, i.e., structures with a low DFT energy can be predicted with very small prediction errors in the pair functional MLIP.
These results indicate that the pair functional and cluster functional MLIPs with AFS structure features can be highly useful for a limited range of applications, although an MLIP with high-order polynomial invariants shows the highest predictive power in general.

\section{Conclusion}
In this study, SO(3) polynomial invariants up to the sixth order representing atomic distributions have been enumerated.
We have applied them to construct accurate MLIPs for elemental Al by formulating the atomic energy as a linear polynomial form of the polynomial invariants.
The high-order invariants play an essential role in constructing an MLIP with a high predictive power for a wide range of crystal structures.
The list of invariants and the group-theoretical procedure to derive the invariants should be useful for constructing not only MLIPs but also prediction models for the other physical properties that depend on the crystal structure.

\begin{acknowledgments}
This study was supported by PRESTO from JST. 
\end{acknowledgments}

\clearpage

\appendix
\section{Projector matrix elements}
\label{sec-appendix-projector}
In this section, we show the formulation used to compute the projector matrix elements for the decomposition of the Kronecker product of Irreps into the identity Irrep.
The projector matrix elements for SO(3) invariants are calculated from integrals involving the Wigner $D$-functions.
The projector matrix elements for the second- and third-order invariants are expressed\cite{AngularMomentumVMK} as
\begin{widetext}
\begin{eqnarray}
^{(1)}P^{l_1l_2}_{m_1m_2,m_1'm_2'} &=&
\frac{1}{8\pi^2}\int_0^{2\pi} d\alpha \int_0^{\pi} d\beta \sin \beta \int_0^{2\pi} d\gamma 
D^{(l_1)}_{m_1m_1'} (\alpha,\beta,\gamma) 
D^{(l_2)}_{m_2m_2'} (\alpha,\beta,\gamma) \nonumber \\
&=&
(-1)^{m_2-m_2'} \frac{1}{2l_2+1} \delta_{l_1l_2} \delta_{-m_1m_2} \delta_{-m_1'm_2'}
\end{eqnarray}
\begin{eqnarray}
^{(1)}P^{l_1l_2l_3}_{m_1m_2m_3,m_1'm_2'm_3'} &=&
\frac{1}{8\pi^2}\int_0^{2\pi} d\alpha \int_0^{\pi} d\beta \sin \beta \int_0^{2\pi} d\gamma 
D^{(l_1)}_{m_1m_1'} (\alpha,\beta,\gamma) 
D^{(l_2)}_{m_2m_2'} (\alpha,\beta,\gamma) 
D^{(l_3)}_{m_3m_3'} (\alpha,\beta,\gamma) \nonumber \\
&=&
(-1)^{m_3-m_3'} \frac{1}{2l_3+1} 
C_{l_1m_1l_2m_2}^{l_3-m_3}
C_{l_1m_1'l_2m_2'}^{l_3-m_3'}
\end{eqnarray}
\end{widetext}
where $C_{l_1m_1l_2m_2}^{l_3m_3}$ denotes the Clebsch--Gordan coefficient for the SO(3) group.
The projector matrix elements for the fourth-, fifth- and sixth-order invariants are reduced to integrals involving products of two or three Wigner $D$-functions using the Clebsch--Gordan expansion given as 
\begin{equation}
\begin{split}
D^{(l_1)}_{m_1m_1'} (\alpha,\beta,\gamma) 
D^{(l_2)}_{m_2m_2'} (\alpha,\beta,\gamma) = \qquad\qquad\qquad \\
\sum_{l=|l_1-l_2|}^{l_1+l_2} \sum_{mm'}
C_{l_1m_1l_2m_2}^{lm} 
D^{(l)}_{mm'} (\alpha,\beta,\gamma) 
C_{l_1m_1'l_2m_2'}^{lm'}. 
\end{split}
\end{equation}
Therefore, the projector matrix elements for the fourth-, fifth- and sixth-order invariants are derived as
\begin{widetext}
\begin{eqnarray}
&& ^{(1)}P^{l_1l_2l_3l_4}_{m_1m_2m_3m_4,m_1'm_2'm_3'm_4'} \nonumber \\
&=&
\frac{1}{8\pi^2}\int_0^{2\pi} d\alpha \int_0^{\pi} d\beta \sin \beta \int_0^{2\pi} d\gamma 
D^{(l_1)}_{m_1m_1'} 
D^{(l_2)}_{m_2m_2'} 
D^{(l_3)}_{m_3m_3'} 
D^{(l_4)}_{m_4m_4'} 
\nonumber \\
&=&
(-1)^{m_4-m_4'} \frac{1}{2l_4+1} 
\sum_{l=|l_1-l_2|}^{l_1+l_2}
C_{l_1m_1l_2m_2}^{l(m_1+m_2)}
C_{l_1m_1'l_2m_2'}^{l(m_1'+m_2')}
C_{l_3m_3l(m_1+m_2)}^{l_4-m_4}
C_{l_3m_3'l(m_1'+m_2')}^{l_4-m_4'}
\end{eqnarray}
\begin{eqnarray}
&& ^{(1)}P^{l_1l_2l_3l_4l_5}_{m_1m_2m_3m_4m_5,m_1'm_2'm_3'm_4'm_5'}  \nonumber \\
&=&
\frac{1}{8\pi^2}\int_0^{2\pi} d\alpha \int_0^{\pi} d\beta \sin \beta \int_0^{2\pi} d\gamma 
D^{(l_1)}_{m_1m_1'} 
D^{(l_2)}_{m_2m_2'} 
D^{(l_3)}_{m_3m_3'} 
D^{(l_4)}_{m_4m_4'} 
D^{(l_5)}_{m_5m_5'} 
\nonumber \\
&=&
(-1)^{m_5-m_5'} \frac{1}{2l_5+1} 
\sum_{l=|l_1-l_2|}^{l_1+l_2}
\sum_{L=|l_3-l|}^{l_3+l}
C_{l_1m_1l_2m_2}^{l(m_1+m_2)}
C_{l_1m_1'l_2m_2'}^{l(m_1'+m_2')}
C_{l_3m_3l(m_1+m_2)}^{L(m_1+m_2+m_3)} \nonumber\\
& & \qquad\qquad\qquad\qquad\qquad
C_{l_3m_3'l(m_1'+m_2')}^{L(m_1'+m_2'+m_3')}
C_{l_4m_4L(m_1+m_2+m_3)}^{l_5-m_5}
C_{l_4m_4'L(m_1'+m_2'+m_3')}^{l_5-m_5'}
\end{eqnarray}
\begin{eqnarray}
&& ^{(1)}P^{l_1l_2l_3l_4l_5l_6}_{m_1m_2m_3m_4m_5m_6,m_1'm_2'm_3'm_4'm_5'm_6'}  \nonumber \\
&=&
\frac{1}{8\pi^2}\int_0^{2\pi} d\alpha \int_0^{\pi} d\beta \sin \beta \int_0^{2\pi} d\gamma 
D^{(l_1)}_{m_1m_1'} 
D^{(l_2)}_{m_2m_2'} 
D^{(l_3)}_{m_3m_3'} 
D^{(l_4)}_{m_4m_4'} 
D^{(l_5)}_{m_5m_5'} 
D^{(l_6)}_{m_6m_6'} 
\nonumber \\
&=&
(-1)^{m_6-m_6'} \frac{1}{2l_6+1} 
\sum_{l=|l_1-l_2|}^{l_1+l_2}
\sum_{L=|l_3-l|}^{l_3+l}
\sum_{S=|l_4-L|}^{l_4+L}
C_{l_1m_1l_2m_2}^{l(m_1+m_2)}
C_{l_1m_1'l_2m_2'}^{l(m_1'+m_2')}
C_{l_3m_3l(m_1+m_2)}^{L(m_1+m_2+m_3)}
C_{l_3m_3'l(m_1'+m_2')}^{L(m_1'+m_2'+m_3')} \nonumber \\
& & \qquad\qquad\qquad
C_{l_4m_4L(m_1+m_2+m_3)}^{S(m_1+m_2+m_3+m_4)}
C_{l_4m_4'L(m_1'+m_2'+m_3')}^{S(m_1'+m_2'+m_3'+m_4')}
C_{l_5m_5S(m_1+m_2+m_3+m_4)}^{l_6-m_6}
C_{l_5m_5'S(m_1'+m_2'+m_3'+m_4')}^{l_6-m_6'}.
\end{eqnarray}
\end{widetext}

\section{List of structure generators}
\label{appendix-structure-generator}

Table \ref{mlip-gtinv-2018:structure-generators} shows the structure generators used for the training and test datasets in this study.

\section{Forces acting on atoms}
\label{appendix-force}
In this section, we derive the formulation for the forces acting on atoms from the atomic energy model with high-order polynomial invariants of Eqn. (\ref{Eqn-linear-polynomial}) used in this study.
The forces acting on atoms are evaluated from the derivatives of the total energy $E$ with respect to the Cartesian coordinates of the atoms.
The Cartesian component $\alpha$ of the force acting on atom $k$, $F_{k,\alpha}$, is written as
\begin{equation}
F_{k,\alpha} = -\frac{\partial E}{\partial x_{k,\alpha}},
\end{equation}
where $x_{k,\alpha}$ denotes the $\alpha$ component of the Cartesian coordinates of atom $k$.
Because the total energy is expressed as the sum of the atomic energies written as Eqn. (\ref{Eqn-linear-polynomial}), the force component is derived as
\begin{equation}
\begin{split}
F_{k,\alpha} = 
- \sum_i 
\left[
\sum_{n} w_{n0} \frac{\partial d_{n0}^{(i)}}{\partial x_{k,\alpha}}
+ \sum_{nl} w_{nll} \frac{\partial d_{nll}^{(i)}}{\partial x_{k,\alpha}} \right. \\
\left. + \sum_{nl_1l_2l_3} w_{nl_1l_2l_3} \frac{\partial d_{nl_1l_2l_3}^{(i)}}{\partial x_{k,\alpha}}
+ \cdots
\right].
\end{split}
\end{equation}
Therefore, the derivatives of polynomial invariants are required to compute the forces.
At the same time, the minus sign of the derivatives of the polynomial invariants corresponds to the structural features for the forces, which are used to estimate regression coefficients as shown in Sec. \ref{sec-regression}.
The derivative on the $p$th-order polynomial invariant is expressed as
\begin{eqnarray}
\frac{\partial d_{nl_1l_2\cdots l_p}^{(i)}}{\partial x_{k,\alpha}} &=& 
\sum_{m_1m_2\cdots m_p} c^{l_1l_2\cdots l_p}_{m_1m_2\cdots m_p} \nonumber\\
&& \Biggl[
\frac{\partial a_{nl_1m_1}^{(i)}}{x_{k,\alpha}} a_{nl_2m_2}^{(i)} \cdots a_{nl_pm_p}^{(i)}\nonumber\\
&+& a_{nl_1m_1}^{(i)} \frac{\partial a_{nl_2m_2}^{(i)}}{\partial x_{k,\alpha}} \cdots a_{nl_pm_p}^{(i)} \nonumber\\
&+& \cdots \nonumber\\
&+& a_{nl_1m_1}^{(i)} a_{nl_2m_2}^{(i)} \cdots \frac{\partial a_{nl_pm_p}^{(i)}}{\partial x_{k,\alpha}} \Biggr],
\end{eqnarray}
where the derivative of $a_{nlm}^{(i)}$ is given by
\begin{equation}
\frac{\partial a_{nlm}^{(i)}}{\partial x_{k,\alpha}} = \sum_{j \in \rm {neighbor}} \frac{\partial}{\partial x_{k,\alpha}} 
\left[ f_n(r_{ij}) Y_{lm}^* (\theta_{ij}, \phi_{ij}) \right].
\end{equation}
The derivative of the right side has a nonzero value when $k=j$ or $k=i$.
When $k=j$, the derivative is computed using the following set of equations:
\begin{widetext}
\begin{eqnarray}
\label{eqn-derivative-FnYlm}
\frac{\partial}{\partial r_{j,x}} \left[ f_n(r) Y_{lm} (\theta, \phi) \right] &=& \frac{x_j-x_i}{r} f_n'(r) Y_{lm} (\theta, \phi) + f_n(r) \left[ \frac{\cos\theta \cos\phi}{r} \frac{\partial Y_{lm}(\theta, \phi)}{\partial \theta} - \frac{\sin\phi}{r\sin\theta} \frac{\partial Y_{lm}(\theta, \phi)}{\partial \phi}\right] \\
\frac{\partial}{\partial r_{j,y}} \left[ f_n(r) Y_{lm} (\theta, \phi) \right] &=& \frac{y_j-y_i}{r} f_n'(r) Y_{lm} (\theta, \phi) + f_n(r) \left[ \frac{\cos\theta \sin\phi}{r} \frac{\partial Y_{lm}(\theta, \phi)}{\partial \theta} + \frac{\cos\phi}{r\sin\theta} \frac{\partial Y_{lm}(\theta, \phi)}{\partial \phi}\right] \\
\label{eqn-derivative-FnYlm2}
\frac{\partial}{\partial r_{j,z}} \left[ f_n(r) Y_{lm} (\theta, \phi) \right] &=& \frac{z_j-z_i}{r} f_n'(r) Y_{lm} (\theta, \phi) + f_n(r) \left[ - \frac{\sin\theta}{r} \frac{\partial Y_{lm}(\theta, \phi)}{\partial \theta} \right],
\end{eqnarray}
\end{widetext}
where the derivative of spherical harmonics with respect to $\theta$ and $\phi$ is given by 
\begin{eqnarray}
\frac{\partial Y_{lm}(\theta, \phi)}{\partial \theta} &=& m \cot\theta Y_{lm}(\theta, \phi) \nonumber \\
&+& \sqrt{(l-m)(l+m+1)} e^{-i\phi} Y_{l(m+1)}(\theta, \phi) \nonumber\\
\frac{\partial Y_{lm}(\theta, \phi)}{\partial \phi} &=& im Y_{lm}(\theta, \phi).
\end{eqnarray}
When $k=i$, the derivatives are given as the minus signs of Eqns. (\ref{eqn-derivative-FnYlm})-(\ref{eqn-derivative-FnYlm2}).

\section{Character of symmetrized $p$th-power of an Irrep}
\label{appendix-sym-character}
The $p$th tensor product of an Irrep is reducible into symmetrized and antisymmetrized representations.
The symmetrized $p$th power of Irrep $^{(\mu)}\Gamma$ and its character are denoted as $[^{(\mu)}\Gamma^p(\hat R)]$ and $\chi[^{(\mu)}\Gamma^p(\hat R)]$, respectively, for an arbitrary operation $\hat R$ of a group.
The character of an operator $\hat R$ in the symmetrized $p$th power of Irrep $\mu$ is expressed as
\begin{eqnarray}
\chi \left[{}^{(\mu)}\Gamma^2(\hat R)\right] &=& \frac{1}{2} {}^{(\mu)}\chi^2(\hat R) + \frac{1}{2} {}^{(\mu)}\chi(\hat R^2) \nonumber \\
\chi \left[{}^{(\mu)}\Gamma^3(\hat R)\right] &=& 
\frac{1}{6} {}^{(\mu)}\chi^3(\hat R) 
+ \frac{1}{2} {}^{(\mu)}\chi(\hat R) {}^{(\mu)}\chi(\hat R^2) 
\nonumber \\
&&
+ \frac{1}{3} {}^{(\mu)}\chi(\hat R^3) 
\nonumber \\
\chi \left[{}^{(\mu)}\Gamma^4(\hat R)\right] &=& 
\frac{1}{24} {}^{(\mu)}\chi^4(\hat R) 
+ \frac{1}{4} {}^{(\mu)}\chi^2(\hat R){}^{(\mu)}\chi(\hat R^2) 
\nonumber \\
&&
+ \frac{1}{3} {}^{(\mu)}\chi(\hat R){}^{(\mu)}\chi(\hat R^3) 
+ \frac{1}{8} {}^{(\mu)}\chi^2(\hat R^2)
\nonumber \\
&&
+ \frac{1}{4} {}^{(\mu)}\chi(\hat R^4) 
\nonumber \\
\chi \left[{}^{(\mu)}\Gamma^5(\hat R)\right] &=& 
\frac{1}{120} {}^{(\mu)}\chi^5(\hat R) 
+ \frac{1}{12} {}^{(\mu)}\chi^3(\hat R){}^{(\mu)}\chi(\hat R^2) 
\nonumber \\
&&
+ \frac{1}{8} {}^{(\mu)}\chi(\hat R){}^{(\mu)}\chi^2(\hat R^2) 
\nonumber \\
&&
+ \frac{1}{6} {}^{(\mu)}\chi^2(\hat R){}^{(\mu)}\chi(\hat R^3) 
\nonumber \\
&&
+ \frac{1}{6} {}^{(\mu)}\chi(\hat R^2){}^{(\mu)}\chi(\hat R^3) 
\nonumber \\
&&
+ \frac{1}{4} {}^{(\mu)}\chi(\hat R){}^{(\mu)}\chi(\hat R^4) 
\\
&&
+ \frac{1}{5} {}^{(\mu)}\chi(\hat R^5)
\nonumber \\
\chi \left[{}^{(\mu)}\Gamma^6(\hat R)\right] &=& 
\frac{1}{720} {}^{(\mu)}\chi^6(\hat R) 
+ \frac{1}{48} {}^{(\mu)}\chi^4(\hat R){}^{(\mu)}\chi(\hat R^2) 
\nonumber \\
&&
+ \frac{1}{16} {}^{(\mu)}\chi^2(\hat R){}^{(\mu)}\chi^2(\hat R^2) 
+ \frac{1}{48} {}^{(\mu)}\chi^3(\hat R^2) 
\nonumber \\
&&
+ \frac{1}{18} {}^{(\mu)}\chi^3(\hat R){}^{(\mu)}\chi(\hat R^3) 
\nonumber \\
&&
+ \frac{1}{6} {}^{(\mu)}\chi(\hat R){}^{(\mu)}\chi(\hat R^2){}^{(\mu)}\chi(\hat R^3)
\nonumber \\
&&
+ \frac{1}{18} {}^{(\mu)}\chi^2(\hat R^3)
+ \frac{1}{8} {}^{(\mu)}\chi^2(\hat R){}^{(\mu)}\chi(\hat R^4)
\nonumber \\
&&
+ \frac{1}{8} {}^{(\mu)}\chi(\hat R^2){}^{(\mu)}\chi(\hat R^4)
\nonumber \\
&&
+ \frac{1}{5} {}^{(\mu)}\chi(\hat R){}^{(\mu)}\chi(\hat R^5)
\nonumber \\
&&
+ \frac{1}{6} {}^{(\mu)}\chi(\hat R^6).
\nonumber 
\end{eqnarray}

\begin{table*}[tbp]
\begin{ruledtabular}
\caption{
ICSD IDs and ICSD structure types of structure generators.
}
\label{mlip-gtinv-2018:structure-generators}
\begin{tabular}{cc|cc|cc}
ICSD CollCode & Structure type & ICSD CollCode & Structure type & ICSD CollCode & Structure type\\
\hline
2091  & Se$_3$S$_5$       &76156  & bcc-W         & 187431 & CaHg$_2$  \\
15535 & O$_2$(hR2)        &76166  & U(beta)-CrFe  & 188336 & (Ca8)xCa2  \\
24892 & N$_2$             &88815  & Graphite(oS16)& 189436 & B(tP50)  \\ 
26463 & S$_{12}$          &88820  & Si(HP)        & 189460 & MnP  \\     
26482 & N$_2$(cP8)        &97742  & Sb$_2$Te$_3$  & 189786 & Si(oS16)  \\
27249 & CO                &104296 & Hg(LT)        & 193439 & Graphite(2H) \\
28540 & H$_2$             &105489 & FeB           & 194468 & I$_2$  \\  
30101 & Wurtzite-ZnS(2H)  &108326 & Cr$_3$Si      & 236662 & Sn  \\
30606 & Se(beta)          &108682 & Pr(hP6)       & 236858 & O$_8$  \\      
31170 & C(P63mc)          &109012 & Li(cI16)      & 245950 & Ge(cF136)  \\
31692 & Pu(mP16)          &109018 & Cs(HP)        & 245956 & Ge  \\
31829 & Graphite(3R)      &109027 & Sr(HP)        & 246446 & Bi(III)  \\
37090 & S$_6$             &109028 & Bi(I4/mcm)    & 248474 & Pu(oF8)  \\
41979 & Diamond-C(cF8)    &109035 & Ca$_{0.15}$Sn$_{0.85}$ & 248500 & O$_2$(mS4)  \\
42679 & Sb(mP4)           &157920 & IrV           & 280872 & Ta(tP30)  \\
43058 & Mn(alpha)-Mn(cI58)&161381 & K(HP)         & 609832 & P(black)  \\
43211 & Po(alpha)         &162242 & Ca(III)       & 616526 & As  \\
43216 & Sn(tI2)           &162256 & Ga            & 639810 & In  \\
43251 & S$_8$(Fddd)       &165132 & B$_{28}$        & 642937 & Mn(cP20)  \\
43539 & Ga(Cmcm)          &165725 & Co(tP28)      & 644520 & N$_2$(epsilon) \\
52412 & Sc                &167204 & Ge(hP8)       & 648333 & Pa  \\
52501 & Te(mP4)           &168172 & I(Immm)       & 652633 & Sm  \\
52914 & ccp-Cu            &173517 & Ge(HP)        & 652876 & hcp-Mg  \\
52916 & La                &181082 & C(P1)         & 653045 & Se(gamma)  \\
56503 & GaSb              &181908 & Si(I4/mmm)    & 653381 & U  \\
56897 & SmNiSb            &182587 & Nickeline-NiAs& 653719 & Bi  \\
57192 & Cs(tP8)           &182732 & BN(P1)        & 653797 & Pu(mS34)  \\
62747 & B(hR12)           &182760 & C(C2/m)       & 656457 & Po(hR1)  \\
76041 & U(beta)           &185973 & C$_3$N$_2$  
\end{tabular}
\end{ruledtabular}
\end{table*}

\bibliography{mlip}

\end{document}